\documentclass[
    ,final            % use final for the camera ready runs 
    ,numberedheadings % uncomment this option for numbered sections
    ,cmfonts          % add further options here if necessary
  ]
  {aipproc}

\usepackage{amsmath}
\usepackage{amssymb}
\layoutstyle{6x9}
\usepackage[breaklinks=true, debug=true, colorlinks=true, linkcolor=red, citecolor=blue, urlcolor=blue]{hyperref}

\begin{document}

\title{A look at the influence of the $J/\psi$ transverse momentum on shadowing}


\author{Andry M. Rakotozafindrabe}{
  address={CEA Saclay, IRFU/Service de Physique Nucléaire, 91191 Gif-sur-Yvette, France}
}

\begin{abstract}

Stringent physical constraints relate the $J/\psi$ produced at a given transverse momentum~$p_T$ to the Bjorken-$x$ of the initial gluons. We present a new approach which takes them into account in order to explicitely investigate the $p_T$-dependence of the shadowing effect on the $J/\psi$ production. Using the $J/\psi$ rapidity and $p_T$ spectra extracted from $\sqrt{s}=200\mathrm{~GeV}$ p~+~p data from PHENIX, we build a Glauber Monte-Carlo code which includes shadowing in two alternative ways: multiple scattering corrections or $Q^2$ evolution of parton densities. We present our results in d~+~Au collisions at the same energy, notably providing the first prediction of the $J/\psi$ nuclear modification factor as a function of $p_T$, and compare them to the available data by adding some nuclear absorption effect.
\end{abstract}

\maketitle

%\tableofcontents

%%%%%%%%%%%%%%%%%%%%%%%%%%%%%%%%%%%%%%%%%%%%
%% MAINMATTER
%%%%%%%%%%%%%%%%%%%%%%%%%%%%%%%%%%%%%%%%%%%%

\section{Introduction}

Relativistic heavy ion collisions may be used as a tool to produce the quark-gluon plasma~(QGP), the state of the hadronic matter for extreme temperature and density~\cite{ChengLattice-KarschReviewLattice}. The $J/\psi \, (c\bar{c})$ is foreseen to be a sensitive probe to the QGP formation, due to effects such as dissociation by colour screening in the deconfined medium~\cite{MatsuiSatz}. The PHENIX experiment at RHIC recently measured the $J/\psi$ production in Au~+~Au collisions at $\sqrt{s_{NN}}=200\mathrm{~GeV}$~\cite{jpsiAuAuRun4}. The yield is quite suppressed with respect to p~+~p collisions~\cite{jpsippRun5} scaled by the equivalent number of binary collisions. But the interpretation relies on a proper subtraction of the cold nuclear matter (CNM) effects, known to impact on the $J/\psi$ production in an ordinary nuclear environment -- such as in any p(d)~+~$A$ collisions where the QGP can not be created. $J/\psi$ production in d~+~Au collisions at $\sqrt{s_{NN}}=200\mathrm{~GeV}$ was measured by PHENIX~\cite{jpsidAuRun3new} to establish the experimental baseline. At RHIC, we shall focus on two CNM effects: (i)~the shadowing (initial-state effect) due to the modification of the free nucleon structure function by the nuclear environment, and (ii) the breakup of correlated $c\bar{c}$ pairs (final-state effect) due to multiple scattering with the remaining nucleons from the incident nuclei, often referred to as ``nuclear absorption''.

Intensive theoretical work has been carried out about nuclear effects on the structure functions~\cite{Arneodo}, especially on the shadowing (see the recent review~\cite{nestorReviewShadowing}) and its effects on the $c\bar{c}$ production (see e.\ g.~\cite{CapellaGribovMultScatt,CFshadowing,Vogt}). In this contribution, we present our approach to explicitely investigate the dependence of the shadowing on the $J/\psi$ transverse momentum~$p_T$. Two different shadowing models (CF and EKS) will be used to get the $J/\psi$ production cross-section in nuclei collisions; the latter being described in the framework of a Glauber~\cite{Glauber} Monte-Carlo. We will extend the prior work reported in~\cite{ourPaper} by adding the nuclear absorption in order to be able to compare the obtained CNM effects to PHENIX d~+~Au data, notably providing a very first prediction of their $p_T$-dependence.

\section{The $p_T$-dependence of shadowing models}

\subsection{Shadowing observables}

The free nucleon structure function $F_2^N$ is the sum over the various parton species of their momentum distributions $xf_i(x,Q^2)$ weighted by their charge $e_i$ squared: $F_2^N = \sum_i e_i^2 \,.\ xf_i(x,Q^2)$ where 
$i$ stands for each of the parton species, i.e.\ all valence and sea (anti)quarks, and the gluons,
Bjorken-$x$ is the fraction of the nucleon momentum carried by the parton,  
 $Q^2$ is the energy scale of the process used to probe $F_2^N$,
and $f_i$ is the parton density (PDF). 

Various processes can be used: for~e.\,g.\ electroweak processes like the deep inelastic scattering (DIS)~\cite{Newman:2003vh} $l+N\to l+X$, or the  Drell-Yan (DY) process $\mathrm{p} + N \to \mu^+\mu^-X$, both sensitive to the quarks and antiquarks PDFs. They provide indirect contraints on the gluon PDF -- notably via the deviations of $F_2^N$ from the Bjorken scaling caused by gluon radiation. At RHIC, $J/\psi$ production mainly proceeds through gluon fusion~\cite{LansbergReview}, hence probing the gluon PDF.

The bound nucleon structure function is studied in $l+A$ or p(d)~+~$A$ thanks to the same processes. A summary of the available measurements in the $(x, Q^2)$ plane used to determine the nuclear PDFs (nPDF) can be found in~\cite{dEnterriaSmallx}. At a given $Q^2$, the ratio $R^A_{F_2}(x,Q^2)$ of the bound to the free nucleon structure functions deviates from unity: the shadowing corresponds to the small-$x$ region (usually $x \lesssim 0.1$) where~$R^A_{F_2}<1$, while the anti-shadowing region with~$R^A_{F_2}>1$ lies at intermediate-$x$ (usually $0.1 \lesssim x \lesssim 0.3$). So any process with initial nuclear partons in these regions will be suppressed (resp.\ enhanced). If shadowing is the sole effect on the $J/\psi$ production, then only the gluonic part $R^A_{g}$ of $R^A_{F_2}$ is needed to build the correction factor that relates the cross-section in p(d)~+~$A$ to the one measured in p~+~p:
\begin{equation}
\sigma^{\mathrm{p(d)}A} = R^A_{\mathrm{shadow}} \times \left< N_\mathrm{coll} \right> \sigma^{\mathrm{pp}}
\label{eq:jpsidAcrossSection}
\end{equation}
where $R^A_{\mathrm{shadow}} = f( R^A_{g} )$ and $\left< N_\mathrm{coll} \right>$ is the average number of collisions in~p(d)~+~$A$. Eq.~\eqref{eq:jpsidAcrossSection} is identical for p~+~$A$ and d~+~$A$ since shadowing effects in the deuterium are negligible. Only its spatial extension must be taken into account. 

As can be intuited from Eq.~\eqref{eq:jpsidAcrossSection}, the relevant experimental observable is the $J/\psi$ nuclear modification factor~$R_{\mathrm{d}A}$, defined as the ratio of the production cross-section in d~+~$A$ to the one in p~+~p scaled by $\left< N_\mathrm{coll} \right>$, or equivalently computed as:
\begin{equation}
R_{\mathrm{d}A} = \frac{dN^{J/\psi}_{\mathrm{dA}}/dy}{\left< N_\mathrm{coll} \right> dN^{J/\psi}_{\mathrm{pp}}/dy}
\label{eq:RdA}
\end{equation}
where $dN^{J/\psi}_{\{\mathrm{d}A, \, \mathrm{pp}\}}/dy$ are the measured $J/\psi$ yield per rapidity unit.

\subsection{Physical origin and models of shadowing}

Shadowing appears as the consequence of coherence effects~\cite{Arneodo,nestorReviewShadowing}. 

\paragraph{CF approach}

Let us describe the shadowing in the target nucleus rest frame. In DIS, the incoming virtual photon fluctuates into a $q\bar{q}$~pair long before reaching the nucleus. At high energy (or at low-$x$), its coherence length\footnote{$m_N$ stands for the nucleon mass.} $l_C = 1 / 2 m_N x$ can become of the order of the nuclear radius, leading to a coherent interaction with several nucleons at once. The cross-section per nucleon is then reduced, which gives birth to shadowing. The hadronic component of the virtual photon interacts with a pomeron ``emitted'' by a nucleon, and shadowing is the outcome of the Gribov theory~\cite{nestorReviewShadowing,CapellaGribovMultScatt,GribovGlauberTh-Kaidalov:2007} composed of a multiple-scattering (multi-pomeron exchanges) formalism and a diffraction component from interactions between the pomerons (which essentially are gluons). The CF~shadowing model~\cite{CapellaGribovMultScatt,CFshadowing,ourPaper} belongs to such approaches. Following the Schwimmer unitarization scheme~\cite{Schwimmer:1975bv} to sum diagrams with triple pomeron interaction, the correction factor due to shadowing in p~+~$A$ collisions can be expressed at fixed impact parameter\footnote{$b$ denotes a position in the transverse plane.}~$b$ as:
\begin{equation}
R^A_{\mathrm{shadow}} (b, y, p_T) = R^A_{\mathrm{Sch}} (b, y, p_T) \overset{def}{\equiv} \frac{1}{1 + A T_A(b) F(y, p_T) }
\label{eq:RshadowCF}
\end{equation}
where $y$ is the center of mass rapidity of the produced particle, $p_T$ is its transverse momentum, $T_A(b)$ is the nuclear profile function related to the Woods-Saxon distribution $\rho_A(b, z)$ at $z$ longitudinal location in space by $\int dz \, \rho_A(b,z) = A \, T_A(b)$. The function $F(y, p_T)$ accounts for initial interactions between gluons. It is given by the integral of the ratio of the triple pomeron cross-section to the single pomeron~one:
\begin{equation}
F(y, p_T) = \left. 4 \pi \int_{y_{min}}^{y_{max}} dy \frac{1}{\sigma^{P}} \frac{d^2 \sigma^{PPP}}{dy dt} \right |_{t=0} = C \left [\exp \left (\Delta \,.\ y_{max}\right ) - \exp \left ( \Delta \,.\ y_{min}\right )\right ]
\label{eq:FfunctionCF}
\end{equation}
where $t=(p-p')^2$ is the usual variable in DIS related to the difference between the incoming and outgoing quadri-momentum of the nucleon, $y_{min}=\ln{\left(\tfrac{R_A m_N}{\sqrt{3}}\right)}$ and $y_{max}=\frac{1}{2}\ln{\left(\frac{s}{m_T^2}\right)} \mp y$ with $y>0 \:(y<0)$ for the projectile (target) hemisphere, with $R_A$ being the nuclear radius, $s$ the square of the center-of-mass energy per collision and $m_T$ the transverse mass of the produced meson, $m_T=\sqrt{m_{J/\psi}^2+p_T^2}$. $C$ is a function of the parameter $\Delta$, the pomeron-proton coupling $g^P_\mathrm{pp}(0)$ and the triple pomeron coupling $r^{PPP}(0)$, both evaluated at $t=0$: $C=\frac{g^P_\mathrm{pp}(0)r^{PPP}(0)}{4\Delta}$. The values of $C$ and $\Delta$ can be fixed from data on DIS. We have used $C=0.31\mathrm{~fm}^2$ and $\Delta=0.13$ as in~\cite{CapellaGribovMultScatt} and references therein.

\paragraph{EKS approach}

Equivalently, the physical picture of the (anti-)shadowing can be viewed from the Breit frame (nucleus infinite momentum frame). Low-$x$ partons spread over a large longitudinal distance $\Delta z$, proportional to $1/(m_Nx)$. Partons from different nucleons may spatially overlap, interact and fuse, hence increasing the parton density at higher-$x$ at the expense of the one at low-$x$ (conservation of the nucleon momentum). 
The EKS model makes use of the fact that nPDFs and PDFs are different as a starting point, without addressing the physical origin. This kind of models rather use the experimental data to provide a parametrization\footnote{However, deriving the nPDFs and hence the ratios $R_i^A(x,Q^2)$ from the data is difficult: in the nuclear case, there is an additional dependence in $A$ and $Z$, and there is no DIS data below $x\lesssim 5 \,.10^{-3}$ at $Q^2 \gtrsim \Lambda_{\mathrm{QCD}}^2$. So the nuclear ratio $R_g^A(x,Q^2)$ for the gluons is less constrained at low-$x$, which is accessible by high energy colliders for heavy quark production. Note also that the existing DIS data do not allow to determine the $b$-dependence of the ratios $R_i^A(x,Q^2)$. But we know from PHENIX results that there is a centrality dependence of $R_{\mathrm{dAu}}$, the $J/\psi$ nuclear modification factor.} of all the parton-specie dependent ratios $R_i^A(x,Q^2_0)$ of nPDF/PDF, at some fixed scale $Q_0^2$, which enter into $R^A_{F_2}(x,Q^2_0)$ by a sum. Although perturbative QCD (pQCD) cannot give the absolute $R_i^A(x,Q^2)$, it can predict their evolution with~$Q^2$ starting from some initial $Q_0>\Lambda_{\mathrm{QCD}}$, thanks to the DGLAP~\cite{DGLAP} equations. \\
The EKS model is named after the EKS98 parametrization~\cite{Eskola:1998iy-Eskola:1998df} used to compute the shadowing correction factor. In EKS98, the ratios $R_i^A(x,Q^2)$ are obtained at $Q_0^2 = 2.25\mathrm{~GeV}^2$, are evolved at LO for $Q^2<10^4\mathrm{~GeV}^2$ and are valid for $x \geq 10^{-6}$. In this model, the spatially-dependent shadowing in p~+~A collisions is given by~\cite{Vogt}:
\begin{equation}
R^A_{\mathrm{shadow}} (b, x, Q^2) = R^A_i(b, x, Q^2) \overset{def}{\equiv} 1 + N_{\rho} \frac{ \int dz \, \rho_A(b,z) }{ \int dz \, \rho_A(0,z) }  \left [ R^A_i(x, Q^2) -1 \right ]
\label{eq:RshadowEKS}
\end{equation}
where $\rho_A(b,z)$ is normalized to get $\int d^2b \int dz \,\rho_A(b,z) =A$, $N_{\rho}$ is a normalization factor chosen to have $(1/A) \int d^2b \int dz \, \rho_A(b,z) R_i^A(b,x,Q^2) = R_i^A(x,Q^2)$. There is a simple idea behind Eq.~\eqref{eq:RshadowEKS}. At large~$b$, the nucleon density is small, so we expect the target nucleons to behave as free nucleons: the shadowing effects should be negligible. At $b=0$, we rather probe the center of the target nucleus, where the vicinity of a high density of nucleons should lead to larger effects. In this equation, the integral over~$z$ includes the target nucleus material that the projectile nucleon traveled through, by its longitudinal path at an impact parameter~$b$. So the spatially-dependent shadowing is obtained by assuming that the projectile parton interacts coherently with all the target partons localized within a cylinder, its axis being defined by its longitudinal path and its transverse section area by the nucleon transverse area~$\sigma^N_{\mathrm{tr}}$. The average nucleon density being $\rho_0 = (\frac{4}{3} \pi r^3)^{-1} = 0.17\mathrm{~nucleon/fm}^3$, we get $\sigma^N_{\mathrm{tr}} = \pi r^2 = 3.94\mathrm{~fm}^2$. Therefore, the number of target partons which contribute to shadowing will be larger at small~$b$.  

\subsection{Where is the $p_T$ dependence?}

For the $J/\psi$, the shadowing is nothing more than a production process where the initial partons are picked within the nPDFs, the latter exhibiting a different dependence on $x$ and $Q^2$ with respect to the PDFs. The underlying production process puts stringent physical constraints on the initial partons that can make a $J/\psi$ at some given rapidity~$y$ and transverse momentum~$p_T$, through quadri-momentum conservation and partonic cross-section dependence on $\hat{s} = s_{NN} x_1 x_2$. Hence, specifying the production process comes down to defining the shadowing dependence on~$y$ and~$p_T$. The simplest production process is 
\begin{equation}
g+g \to c\bar{c}
\label{eq:jpsiSimpleProdProcess}
\end{equation}
At $p_T=0$, quadri-momentum conservation results in:
\begin{equation}
x_{1,2} = \frac{m_{J/\psi}}{\sqrt{s_{NN}}}\exp(\pm y)
\label{eq:xUsedByRamona}
\end{equation}
Keeping this simplest process for $p_T \neq 0$ implies that the initial partons cannot be colinear: they have to carry an intrinsic transverse momentum, later on transferred to the created $c\bar{c}$. If the $p_T$ of the $J/\psi$ has such an intrinsic origin, then:
\begin{equation}
x_{1,2} = \frac{m_T}{\sqrt{s_{NN}}}\exp(\pm y)
\label{eq:xIntrinsic}
\end{equation}

The available theoretical predictions of the EKS shadowing~\cite{Vogt} were all made at $p_T=0$ with the use Eq.~\eqref{eq:xUsedByRamona}. In our work~\cite{ourPaper}, we will investigate how these predictions, made at RHIC energy, are affected by the introduction of a non-zero $p_T$ when using Eq.~\eqref{eq:xIntrinsic}. We also add the non-zero~$p_T$ in the energy scale:
\begin{equation}
Q^2 = (2m_c)^2 + (p_T)^2
\label{eq:Q2Intrinsic}
\end{equation}
where $m_c=1.2\mathrm{~GeV}/c^2$ is the charm mass value, in accordance with~\cite{Vogt}.

About the CF shadowing~\cite{CapellaGribovMultScatt}, the explicit $p_T$-dependence of Eq.~\eqref{eq:RshadowCF} rely on the assumption in Eq.~\eqref{eq:xIntrinsic}. The previously published predictions~\cite{CFshadowing} are made at a fixed value of~$y$ and a unique value of $p_T\sim0$ since they only consider $m_T = 3.1\mathrm{~GeV}$. In our work~\cite{ourPaper}, we will allow both $y$ and $p_T$ to vary within their full spectra.

\section{$J/\psi$ production in our Monte-Carlo}

To investigate the CNM effects on the $J/\psi$ production, we implemented a Monte-Carlo framework, with three main steps.

\paragraph{Step 1: nucleus-nucleus collisions described with a Glauber~\cite{Glauber} model} 

For each $A+B$ collision (an event), the value of the impact parameter~$b$ is randomly chosen relative to the so-called ``minimum bias'' distribution ($2 \pi b \, db$). Then, for each nucleus, the nucleon positions are randomly chosen according to the nuclear density profile, defined by the Woods-Saxon~\cite{WoodsSaxon} parametrization for any nucleus $A>2$ and by the Hulthen~\cite{Hulthen} parametrization for the deuterium. Within this code, we can determine the number of target nucleons in the path of each projectile nucleon and calculate an event-by-event $N_{\mathrm{coll}}$. Such a binary collision is considered to occur if the distance~$d$ between two nucleons satifies~$\pi d^2 < \sigma_{NN}$, where $\sigma_{NN}$ stands for the nucleon-nucleon cross-section (at RHIC,  $\sigma_{NN}=42\mathrm{~mb}$ at $\sqrt{s_{NN}}=200\mathrm{~GeV}$). 

\paragraph{Step 2: kinematics for the produced $J/\psi$ candidates}

For each nucleon-nucleon collision, a $J/\psi$ candidate can be produced (with an arbitrary production cross-section $\sigma_{J/\psi} \lesssim \frac{\sigma_{NN}}{2}$), with $y$ and $p_T$ randomly chosen in the respective input spectra. The latter are given by fits to the recent PHENIX p~+~p data~\cite{jpsippRun5} at $\sqrt{s_{NN}}=200\mathrm{~GeV}$. The angular orientation~$\varphi$ of $p_T$ in the $(p_x,p_y)$ plane is also random and uniformly distributed in $[0, 2\pi]$. The Bjorken-$x$ carried by the initial partons are then computed with Eq.~\eqref{eq:xIntrinsic}, and the scale~$Q^2$ with Eq.~\eqref{eq:Q2Intrinsic}. In order to remain in the physical phase space domain, we require that $0 < x_1, x_2 < 1$.

\paragraph{Step 3: involving CNM effects}

To account for shadowing, only some $J/\psi$ candidates are randomly allowed to become ``real'' $J/\psi$, since the production cross-section of the real $J/\psi$ is $\sigma_{J/\psi}$ corrected by the factor $R_{\mathrm{shadow}}^{AB}$, where
\begin{equation}
R_{\mathrm{shadow}}^{AB} = R_{\mathrm{shadow}}^{A} \times R_{\mathrm{shadow}}^{B} 
\label{eq:RabShadow}
\end{equation}
and $R_{\mathrm{shadow}}^{\{A,\,B\}}$ are given by Eq.~\eqref{eq:RshadowCF} and \eqref{eq:RshadowEKS} for the CF and EKS~models, respectively. To account for the nuclear absorption, any Monte-Carlo $J/\psi$ that is ``wounded'' by the remaining incident nucleons is tagged as broken. The cross-section~$\sigma_{\mathrm{break-up}}$ is used to define the distance~$d$ required to get a $J/\psi$-nucleon interaction: $\pi d^2 < \sigma_{\mathrm{break-up}}$. Finally, the $J/\psi$ nuclear modification factor $R_{AB}$ is computed as followed:
\begin{equation}
R_{AB} = \frac{dN_{\mathrm{not~wounded~real~} J/\psi}/dy}{dN_{J/\psi\mathrm{~candidate}}/dy}
\label{eq:MC-Rab}
\end{equation}

\section{Results at RHIC energy and discussion}

%\paragraph{About the influence of $p_T$ on the shadowing}

\begin{figure}[thb]
  \includegraphics[height=7cm]{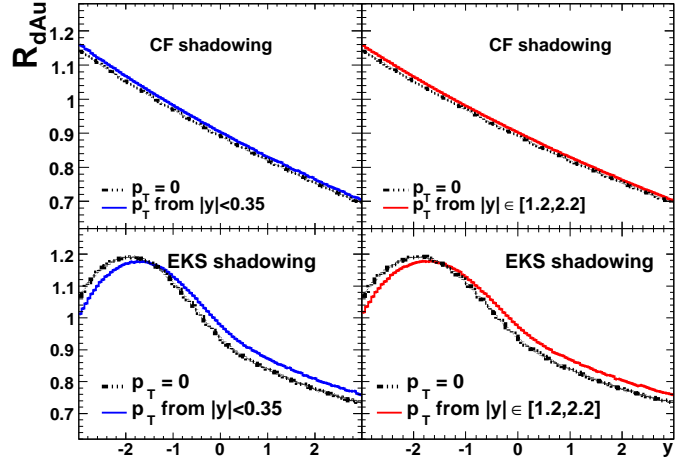}
  \caption{$J/\psi$ nuclear modification factor in d~+~Au collisions as a function of rapidity for CF (top) and EKS (bottom) shadowing models. Three input $p_T$ distributions are used: $p_T = 0$, $p_T$ from $|y|<0.35$ and $p_T$ from $1.2<|y|<2.2$.}
\label{fig:RdAu-vs-y}
\end{figure}

Fig.~\ref{fig:RdAu-vs-y} shows $R_{\mathrm{dAu}}$ as a function of~$y$ when considering or not a $p_T$-dependence of the CF and EKS models. Here, shadowing is the sole effect studied, so $R_{\mathrm{dAu}}$ can be identified to $R^{A}_{\mathrm{shadow}}$. Adding an intrinsinc $p_T$-dependence has a slight effect only (more visible for the EKS case), because of the average value of $p_T$ in Eq.~\eqref{eq:xIntrinsic}: $\left< p_T \right> < 2\mathrm{~GeV/c} < m_{J/\psi}$  at RHIC energy. For CF, $F(y,p_T)$ is a monotonic function, increasing with~$y$ and decreasing with~$p_T$. So at fixed~$y$, a larger~$p_T$ results in a smaller value of $F(y,p_T)$ and hence a larger $R^A_{\mathrm{shadow}}$ from Eq.~\eqref{eq:RshadowCF}. For EKS, a larger~$p_T$ leads to a larger~$Q^2$. In the higher-$x$ part of the anti-shadowing region, this leads to a smaller $R_i^A(x,Q^2)>1$, and hence a smaller $R^A_{\mathrm{shadow}}$ from Eq.~\eqref{eq:RshadowEKS} (the contrary is obtained for the lower-$x$ part of the shadowing region). Fig.~\ref{fig:RdAu-vs-y} also shows that the amount of antishadowing is larger for EKS compared to CF.

\begin{figure}[htb]
  \includegraphics[height=7.5cm]{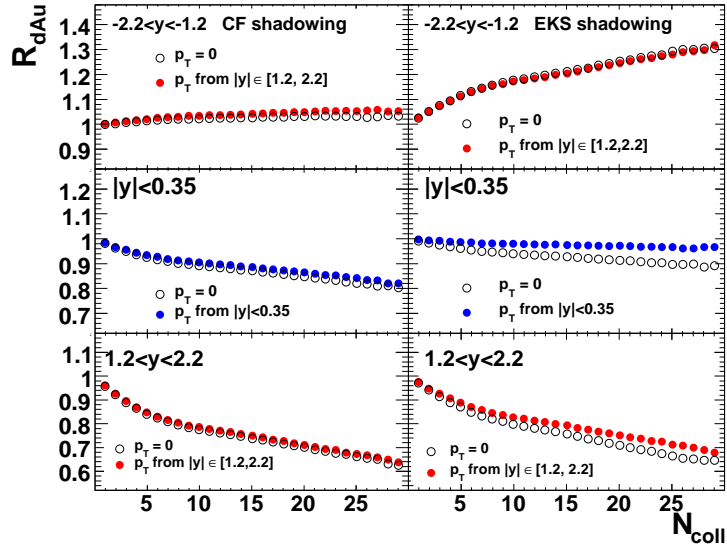}
  \caption{$J/\psi$ nuclear modification factor in d~+~Au collisions as a function of the number of collisions for CF (left) and EKS (right) shadowing models for three rapidity windows: backward $(-2.2<y<-1.2)$, central $(|y|<0.35)$ and forward $(1.2<y<2.2)$ regions (from up to down).}
\label{fig:RdAu-vs-Ncoll}
\end{figure}

Fig.~\ref{fig:RdAu-vs-Ncoll} presents the expected $R_{\mathrm{dAu}}$ as a function of~$\left< N_{\mathrm{coll}} \right>$ for the three rapidity windows accessible by the PHENIX detector. There is a remarkable difference between the effect predicted by CF and EKS models at backward rapidity: the smaller amount of anti-shadowing for the CF model shows up here. This does not come as a surprise, since the CF model was originally built to describe the coherence effect that leads to the depletion of the nuclear structure function at low-$x$. Its validity domain for the present energies starts around $y \gtrsim-2$.

\begin{figure}[htb]
  \includegraphics[height=7cm]{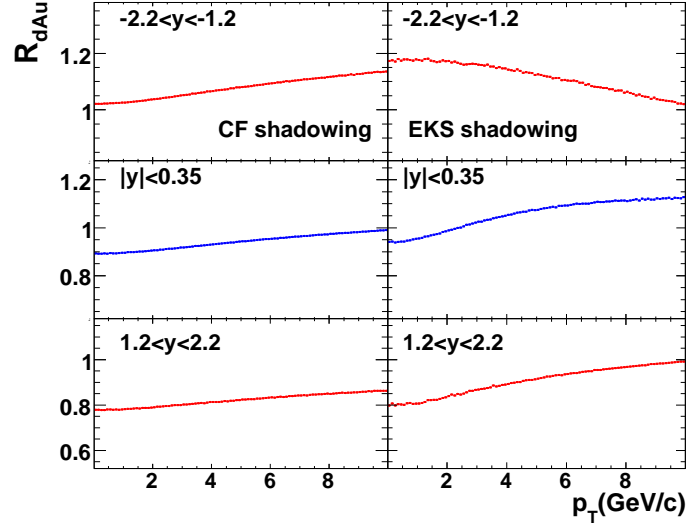}
  \caption{$J/\psi$ nuclear modification factor in d~+~Au collisions as a function of the $J/\psi$ transverse momentum for CF (left) and EKS (right) shadowing for the backward, central and forward rapidity regions (from up to down).}
\label{fig:RdAu-vs-pT}
\end{figure}

Our main result lies in Fig.~\ref{fig:RdAu-vs-pT}, which shows the expected $R_{\mathrm{dAu}}$ as a function of $p_T$ for both models. The $p_T$-dependence is rather significant: it can lead to amplitude variations as large as about $20\%$. The EKS model exhibits a stronger dependence with~$p_T$. At backward rapidity, EKS et CF notably show opposite behaviours. 

%\paragraph{Comparison to the data}
%\vspace*{-0.8cm}
\begin{figure}[htb]
  \includegraphics[height=8cm]{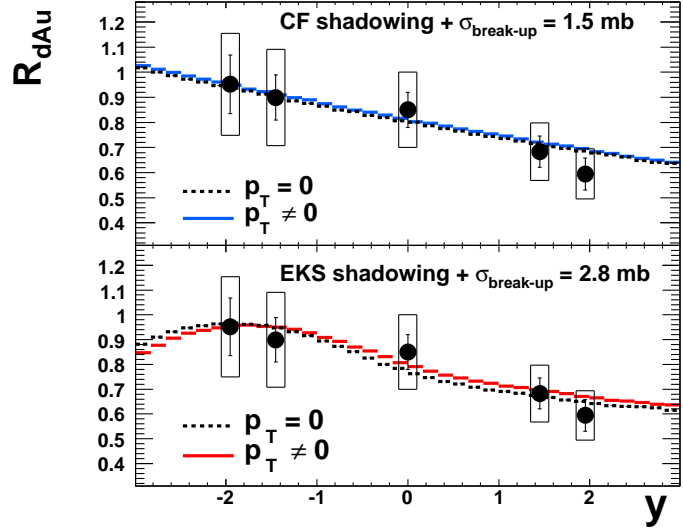}
  \caption{$J/\psi$ nuclear modification factor in d~+~Au collisions as a function of rapidity for CF (top) and EKS (bottom) shadowing on top of the nuclear absorption and compared to PHENIX data~\cite{jpsidAuRun3new}. Bars stand for point to point uncorrelated errors (both statistical and systematical uncertainties) and boxes for the point to point correlated systematical errors.}
\label{fig:dataRdAu-vs-y}
\end{figure}

\begin{figure}[htb]
  \includegraphics[height=7.5cm]{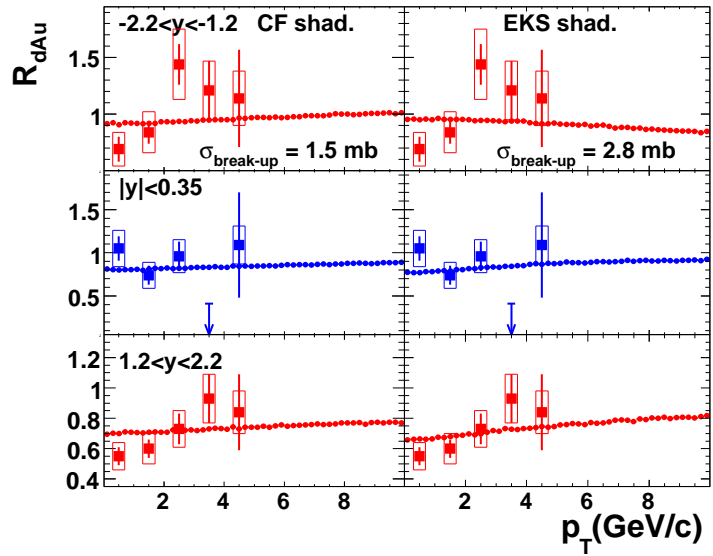}
  \caption{$J/\psi$ nuclear modification factor in d~+~Au collisions as a function of the $J/\psi$ transverse momentum for CF (left) and EKS (right) shadowing on top of the nuclear absorption and compared to PHENIX data~\cite{jpsidAuRun3new} for the three rapidity regions. Bars stand for point to point uncorrelated errors and boxes for the point to point correlated errors.}
\label{fig:dataRdAu-vs-pT}
\end{figure}

The predictions with shadowing as the sole effect overshoot the data, which advocates for an additional nuclear absorption effect. It is parametrized by the break-up cross-section $\sigma_{\mathrm{break-up}}$, not calculable from first QCD principles. For each shadowing model, we choose $\sigma_{\mathrm{break-up}}$ in order to get the best description of $R_{\mathrm{dAu}}$ vs $y$ as measured by PHENIX~\cite{jpsidAuRun3new} . The results are shown on Fig.~\ref{fig:dataRdAu-vs-y}: different values of the cross-section are required for the CF model ($\sigma_{\mathrm{break-up}} = 1.5\mathrm{~mb}$) and the EKS model ($\sigma_{\mathrm{break-up}} = 2.8\mathrm{~mb}$ i.\ e.\ the same value as in~\cite{jpsidAuRun3new}). The corresponding expectation of $R_{\mathrm{dAu}}$ as a function of $p_T$ is compared to PHENIX data~\cite{jpsidAuRun3new} on Fig.~\ref{fig:dataRdAu-vs-pT}. Unfortunately, their large uncertainties and limited range in~$p_T$ do not allow to draw firm conclusions. At backward rapidity, their trend with~$p_T$ seems to disagree with the EKS model. For the other rapidity regions, both models give an approximate description of the data. At forward rapidity, they both seem to underestimate the slope, which may indicate some missing effect (like a possible broadening of the transverse momemtum of the initial gluons due to the multiple-scattering experienced as the incident nucleon travels through the nucleus).

\vspace*{-0.5cm}
\section{Conclusions and outlook}

In summary, we compared the $J/\psi$ shadowing expected from two models, CF and EKS, in d~+~Au collisions at RHIC described in the framework of a Glauber Monte-Carlo. In general, the suppresion is stronger for CF, which implies a smaller break-up cross-section for the nuclear absorption effect needed to match the data. We investigated their $p_T$-dependence based on the assumption that the $J/\psi$'s $p_T$ would be of intrinsic origin and we used the $y$ and $p_T$ spectra from the p~+~p data as an input of the Monte-Carlo. Although the resulting $R_{\mathrm{dAu}}$ vs $y$ or vs $\left< N_{\mathrm{coll}}\right>$ was only slightly affected when introducing such dependence, this approach lead to the first predictions of $R_{\mathrm{dAu}}$ as a function of $p_T$. Due to large uncertainties and a limited range in~$p_T$, the available data do not allow to state definitive conclusions. However, the data seem to have an increasing trend with $p_T$ at backward rapidity, which sound to be in accordance with the expectations from CF rather than EKS. Hopefully, there is much higher statistics d~+~Au data recently taken at RHIC.

For the models, there are two major sources of uncertainty: the parametrization of the ratios nPDFs/PDFs used in EKS-type models (see~\cite{deFlorian:2003qf,Eskola:2008ca}), and the way the $J/\psi$ acquires its $p_T$. The production process $g+g\rightarrow c\bar{c} +g$, with strictly coli\-near initial gluons, can also give a non-zero $p_T$ (extrinsic origin). The corresponding  $p_T$-differential cross-section from~\cite{Haberzettl-Lansberg} reasonably matches both RHIC and Tevatron data. The next step will be to use this computation to relate $J/\psi$'s $(y, p_T)$ to $(x_1,x_2)$ in the extrinsic scheme. The main advantage will be to free the Monte-Carlo from any input spectra from the p~+~p data in order to get predictions at LHC energies.

%%%%%%%%%%%%%%%%%%%%%%%%%%%%%%%%%%%%%%%%%%%%%%%%
%% BACKMATTER
%%%%%%%%%%%%%%%%%%%%%%%%%%%%%%%%%%%%%%%%%%%%%%%%

\begin{theacknowledgments}
This work results from a close collaboration with E.\ Ferreiro and F.~Fleuret. It was presented on behalf of three of us. We are thankful to J.~P.~Lansberg for fruitful discussions. We would like to thank the organizers of the conference.
\end{theacknowledgments}

\bibliographystyle{unsrt}

\begin{thebibliography}{11}

\bibitem{ChengLattice-KarschReviewLattice}
  M.~Cheng {\it et al.},
  %``The transition temperature in QCD'',
  Phys.\ Rev.\  D {\bf 74} (2006) 054507
  \href{http://arxiv.org/abs/hep-lat/0608013}{[hep-lat/0608013]}.
  %%CITATION = PHRVA,D74,054507;%%
 F.~Karsch and E.~Laermann,
  %``Thermodynamics and in-medium hadron properties from lattice QCD'',
in \emph{Quark-Gluon Plasma}, ed. R. Hwa {\it et al.} (2003) 1--59 \href{http://arxiv.org/abs/hep-lat/0305025}{[hep-lat/0305025]}.
  %%CITATION = HEP-LAT/0305025;%%

\bibitem{MatsuiSatz}
  T.~Matsui and H.~Satz,
  %``J/psi Suppression by Quark-Gluon Plasma Formation'',
  Phys.\ Lett.\  B {\bf 178} (1986) 416.
  %%CITATION = PHLTA,B178,416;%%

\bibitem{jpsiAuAuRun4}
  A.~Adare {\it et al.}  [PHENIX],
  %``J/psi production vs centrality, transverse momentum, and rapidity in Au~+~Au collisions at s(NN)**(1/2) = 200-GeV'',
  Phys.\ Rev.\ Lett.\  {\bf 98} (2007) 232301 \href{http://arxiv.org/abs/nucl-ex/0611020}{[nucl-ex/0611020]}.
  %%CITATION = PRLTA,98,232301;%%

\bibitem{jpsippRun5}
  A.~Adare {\it et al.}  [PHENIX],
  %``J/psi production vs transverse momentum and rapidity in p + p collisions at s**(1/2) = 200-GeV'',
  Phys.\ Rev.\ Lett.\  {\bf 98} (2007) 232002
\href{http://arxiv.org/abs/hep-ex/0611020}{[hep-ex/0611020]}.
  %%CITATION = PRLTA,98,232002;%%

\bibitem{jpsidAuRun3new}
  A.~Adare {\it et al.}  [PHENIX],
  %``Cold Nuclear Matter Effects on J/Psi as Constrained by Deuteron-Gold Measurements at sqrt(s_NN) = 200 GeV'',
  Phys.\ Rev.\  C {\bf 77} (2008) 024912 
\href{http://arxiv.org/abs/0711.3917}{[arXiv:0711.3917]}.
  %%CITATION = PHRVA,C77,024912;%%

\bibitem{Arneodo}
  M.~Arneodo,
  %``Nuclear effects in structure functions'',
  Phys.\ Rept.\  {\bf 240} (1994) 301.
  %%CITATION = PRPLC,240,301;%%

\bibitem{nestorReviewShadowing}
  N.~Armesto,
  %``Nuclear shadowing'',
  J.\ Phys.\ G {\bf 32} (2006) R367
\href{http://arxiv.org/abs/hep-ph/0604108}{[hep-ph/0604108]}.
  %%CITATION = JPHGB,G32,R367;%%

\bibitem{CapellaGribovMultScatt}
  N.~Armesto, A.~Capella, A.~B.~Kaidalov, J.~Lopez-Albacete and C.~.A.~Salgado,
  %``Nuclear structure functions at small-x from inelastic shadowing and diffraction'',
  Eur.\ Phys.~J.\ C~{\bf 29} (2003) 531
\href{http://arxiv.org/abs/hep-ph/0304119}{[hep-ph/0304119]}.
  %%CITATION = EPHJA,C29,531;%%

\bibitem{CFshadowing}
  A.~Capella and E.~G.~Ferreiro,
  %``J/psi suppression at s**(1/2) = 200-GeV in the comovers interaction model'',
  Eur.\ Phys.\ J.\  C~{\bf 42} (2005) 419
\href{http://arxiv.org/abs/hep-ph/0505032}{[hep-ph/0505032]}.
  %%CITATION = EPHJA,C42,419;%%

\bibitem{Vogt}
  R.~Vogt,
  %``Inhomogeneous shadowing effects on J/psi production in {\it dA} collisions'',
  Phys.\ Rev.\ Lett.\ {\bf 91} (2003) 142301
  \href{http://arxiv.org/abs/nucl-th/0305046}{[nucl-th/0305046]}.
  %%CITATION = PRLTA,91,142301;%%
  R.~Vogt,
  %``Shadowing and absorption effects on J/psi production in d A collisions'',
  Phys.\ Rev.\ C {\bf 71} (2005) 054902
  \href{http://arxiv.org/abs/hep-ph/0411378}{[hep-ph/0411378]}.
  %%CITATION = PHRVA,C71,054902;%%
  R.~Vogt,
  %``Baseline cold matter effects on J/psi production in A A collisions'',
  Acta Phys.\ Hung.\  A {\bf 25} (2006) 97
\href{http://arxiv.org/abs/nucl-th/0507027}{[nucl-th/0507027]}.
  %%CITATION = APHUE,A25,97;%%

\bibitem{Glauber}
  R.~J.~Glauber,
  in \emph{Lectures in Theoretical Physics},
  ed.\ WE Brittin and LG Dunham, Interscience Publishers (New York), 1  (1959) 315.

\bibitem{ourPaper}
  E.~G.~Ferreiro, F.~Fleuret and A.~Rakotozafindrabe,
  %``Transverse momentum dependence of J/psi shadowing effects'',
\href{http://arxiv.org/abs/0801.4949}{arXiv:0801.4949}.
  %%CITATION = ARXIV:0801.4949;%%

\bibitem{Newman:2003vh}
  P.~Newman,
  %``Deep inelastic lepton-nucleon scattering at HERA'',
  Int.\ J.\ Mod.\ Phys.\ A {\bf 19} (2004) 1061,
  \href{http://arxiv.org/abs/hep-ex/0312018}{[hep-ex/0312018]}.
  %%CITATION = IMPAE,A19,1061;%%

\bibitem{LansbergReview}
  J.~P.~Lansberg,
  %``J/psi, psi' and Upsilon production at hadron colliders: A review'',
  Int.\ J.\ Mod.\ Phys.\  A {\bf 21} (2006) 3857
\href{http://arxiv.org/abs/hep-ph/0602091}{[hep-ph/0602091]}.
  %%CITATION = IMPAE,A21,3857;%%

\bibitem{dEnterriaSmallx}
  D.~G.~d'Enterria,
  % ``Low-x QCD physics from RHIC and HERA to the LHC'',
  Eur.\ Phys.\ J.\ A {\bf 31} (2007) 816,
\href{http://arxiv.org/abs/hep-ex/0610061}{[hep-ex/0610061]}.
  %%CITATION = EPHJA,A31,816;%%

\bibitem{GribovGlauberTh-Kaidalov:2007}
  A.~Capella, A.~Kaidalov, and J.\ Tran Thanh Van,
  %``Gribov theory of nuclear interactions and particle densities at future heavy-ion colliders'',
 Heavy Ion Phys.\ ~{\bf 9} (1999) 169
\href{http://arxiv.org/abs/hep-ph/9903244}{[hep-ph/9903244]}.
  %%CITATION = APHPF,9,169;%%
  K.\ Tywoniuk, I.\ C.\ Arsene, L.\ Bravina, A.\ B.\ Kaidalov and E.~Zabrodin,
%``Gluon shadowing in the Glauber-Gribov model at HERA''
Phys.\ Lett.\ B {\bf 657} (2007) 170,
\href{http://arxiv.org/abs/0705.1596}{[arXiv:0705.1596]}.
%%CITATION = PHLTA,B657,170;%%

\bibitem{Schwimmer:1975bv}
  A.~Schwimmer,
  %``Inelastic Rescattering And High-Energy Reactions On Nuclei'',
  Nucl.\ Phys.\  B {\bf 94} (1975) 445.
  %%CITATION = NUPHA,B94,445;%%

\bibitem{DGLAP}
  Y.\ L.\ Dokshitzer,
  %``Calculation of the structure functions for deep inelastic scattering and {\it e}+ {\it e}- annihilation by perturbation theory in Quantum Chromodynamics. (In Russian)'',
  Sov.\ Phys.\ JETP {\bf 46} (1977) 641.
  %[Zh.\ Eksp.\ Teor.\ Fiz.\ {\bf 73} (1977) 1216].
  %%CITATION = ZETFA,73,1216;%%
  V.\ N.\ Gribov and L.\ N.\ Lipatov,
  %``Deep inelastic {\it e}~+~p scattering in perturbation theory'',
  Sov.\ J.\ Nucl.\ Phys.\ {\bf 15} (1972) 438.
  %[Yad.\ Fiz.\ {\bf 15} (1972) 781].
  %%CITATION = YAFIA,15,781;%%
  G.\ Altarelli and G.\ Parisi,
  %``Asymptotic freedom in parton language'',
  Nucl.\ Phys.\ B {\bf 126} (1977) 298.
  %%CITATION = NUPHA,B126,298;%%

\bibitem{Eskola:1998iy-Eskola:1998df}
  K.~J.~Eskola, V.~J.~Kolhinen and P.~V.~Ruuskanen,
  %``Scale evolution of nuclear parton distributions'',
  Nucl.\ Phys.\  B {\bf 535} (1998) 351
\href{http://arxiv.org/abs/hep-ph/9802350}{[hep-ph/9802350]}.
  %%CITATION = NUPHA,B535,351;%%
  K.~J.~Eskola, V.~J.~Kolhinen and C.~A.~Salgado,
  %``The scale dependent nuclear effects in parton distributions for  practical applications'',
  Eur.\ Phys.\ J.\  C {\bf 9} (1999) 61
\href{http://arxiv.org/abs/hep-ph/9807297}{[hep-ph/9807297]}.
  %%CITATION = EPHJA,C9,61;%%

\bibitem{WoodsSaxon}
  R.\ D.\ Woods and D.\ S.\ Saxon,
  %``Diffuse Surface Optical Model for Nucleon-Nuclei Scattering'',
  Phys.\ Rev.\ {\bf 95} (1954) 577.

\bibitem{Hulthen}
  L.\ Hulth\'en and M.\ Sugarawa,
  %``The two-nucleon problem'',
  Handbook of Physics {\bf 39}, ed.\ Springer-Verlag (1957).

\bibitem{deFlorian:2003qf}
  D.~de Florian and R.~Sassot,
  %``Nuclear parton distributions at next to leading order'',
  Phys.\ Rev.\  D {\bf 69} (2004) 074028
\href{http://arxiv.org/abs/hep-ph/0311227}{[hep-ph/0311227]}.
  %%CITATION = PHRVA,D69,074028;%%

\bibitem{Eskola:2008ca}
  K.~J.~Eskola, H.~Paukkunen and C.~A.~Salgado,
  %``An improved global analysis of nuclear parton distribution functions including RHIC data'',
\href{http://arxiv.org/abs/0802.0139}{arXiv:0802.0139}.
  %%CITATION = ARXIV:0802.0139;%%

\bibitem{Haberzettl-Lansberg}
  J.~P.~Lansberg, J.~R.~Cudell and Yu.~L.~Kalinovsky,
  %``New contributions to heavy quarkonium production'',
  Phys.\ Lett.\  B {\bf 633} (2006) 301
\href{http://arxiv.org/abs/hep-ph/0507060}{[hep-ph/0507060]}.
  %%CITATION = PHLTA,B633,301;%%
  H.~Haberzettl and J.~P.~Lansberg,
  %``Possible solution of the J/psi production puzzle'',
  Phys.\ Rev.\ Lett.\  {\bf 100} (2008) 032006
\href{http://arxiv.org/abs/0709.3471}{[arXiv:0709.3471]}.
  %%CITATION = PRLTA,100,032006;%%
  J.~P.~Lansberg,
  %``Possible solution of the J/psi production puzzle'',
  these proceedings.

\end{thebibliography}

\end{document}